%% file: manuscript-ieee.tex
\documentclass[journal]{IEEEtran}

\usepackage[T1]{fontenc}
\usepackage[utf8]{inputenc}
\usepackage{booktabs,array,graphicx,xcolor,calc,etoolbox}
\usepackage{amssymb}
\usepackage{url}
\usepackage[hidelinks]{hyperref}

\providecommand{\tightlist}{\setlength{\itemsep}{0pt}\setlength{\parskip}{0pt}}

\providecommand{\real}[1]{#1}

\begin{document}

\title{Classical Acceptance Is Not Hybrid Authentication:
Validation Policy and Lifecycle Management of Hybrid X.509 Certificates
in Deployed Open-Source Stacks}

\author{Taesung~Kim,
        Boheung~Chung,
        Keonwoo~Kim,
        and~Yousung~Kang
\thanks{This work has been submitted to the IEEE for possible
publication. Copyright may be transferred without notice, after which
this version may no longer be accessible.}%
\thanks{This work was supported by the Institute of Information and
Communications Technology Planning and Evaluation (IITP) grant funded by
the Korea government (MSIT) (No. RS-2025-02306395, Development and
Demonstration of PQC-Based Joint Certificate PKI Technology; and No.
RS-2026-25519543, Development of PQC Migration Technology for
Quantum-Safe IC Chip). The funders had no role in the study design, data
collection, analysis or interpretation, manuscript preparation, or the
decision to submit the article for publication.}%
\thanks{The authors are with the Electronics and Telecommunications
Research Institute (ETRI), 218 Gajeong-ro, Yuseong-gu, Daejeon 34129,
Republic of Korea (e-mail: taesung@etri.re.kr).}}

\markboth{IEEE Transactions on Network and Service Management}%
{Kim \MakeLowercase{\textit{et al.}}: Classical Acceptance Is Not Hybrid Authentication}

\IEEEtitleabstractindextext{%
\begin{abstract}
\input{abstract.tex}
\end{abstract}

}

\maketitle
\IEEEdisplaynontitleabstractindextext
\IEEEpeerreviewmaketitle

\input{sections.tex}

\section*{Data Availability}
The reproducibility package contains the deterministic corpus generator,
a manifest with per-artifact DER SHA-256 digests, the stack adapters,
the annotated reference procedure, the generator and fixed vectors for
the invalidated variants, and the recorded observations with raw logs.
It is archived at Zenodo (doi:10.5281/zenodo.21452130) and developed at
\url{https://github.com/taesung901-ui/pqt-verifier-semantics-artifact}.
All certificates are synthetic, issued under a dedicated test hierarchy;
no real certificates, private keys, or personal data are included.

\input{refs.tex}
\end{document}

%% file: abstract.tex
Post-quantum migration relies on hybrid X.509 certificates, which carry
post-quantum material alongside the classical so existing verifiers
still work. Several designs place it where a verifier may ignore it, so
the classical path decides. We tested eight open-source
path-validation stacks over seven independent codebases, one in two
builds: nine configurations over six certificate profiles. On their
default paths, every stack that parsed a separable hybrid certificate
accepted it. Invalidating the post-quantum evidence in each separable
scheme, leaving the classical evidence valid, changed no verdict in any
of the 27 cells: none distinguished sound post-quantum evidence from
destroyed. Four stacks verify post-quantum signatures elsewhere on the
same path, so immature support does not explain it. Two stacks implement
the checks the schemes specify, neither on its default path, and no
document defines the interface between them: one carrying a relying
party's hybrid requirement, an operational policy, into path validation
and reporting which kind of acceptance resulted. We contribute a
specification-derived model, this test, and a policy-parametric contract
pairing a policy input with a labelled result. Revoking a bound
post-quantum certificate changes no verdict in any of the nine
configurations, because none consults it; the labelled result makes it
visible to operations.

%% file: sections.tex
\section{Introduction}\label{sec:1}
Organizations are beginning to replace the RSA and elliptic-curve
signatures in their public-key infrastructure with post-quantum
algorithms, because a future quantum computer would make those
signatures forgeable. A certificate carrying only a post-quantum
signature is rejected by verifiers already deployed across the Internet,
which do not implement the new algorithms. The transitional answer,
specified in several competing designs, is the \emph{hybrid
certificate}, which carries post-quantum key and signature material
alongside the classical material: an old verifier still validates the
classical part, and a new one can also check the post-quantum part.
Acceptance by the old verifier is the design goal of these schemes
rather than an accident; Bindel et al.~demonstrated it for X.509 in 2017
\cite{r19}. The prior literature has not examined the other side: what
post-quantum-capable verifiers check by default, what happens when the
bound material's lifecycle diverges, and what the validation interface
reports about either. Lee et al.'s concurrent work \cite{r18} addresses
binding enforcement, but not the lifecycle or interface questions.

A \emph{verifier} is the software component inside a TLS library or PKI
toolkit that performs certificate path validation and returns accept or
reject. A \emph{relying party} is the entity that depends on that result
to make a security decision, under a policy it has stated. The
distinction matters because the policy and the code that serves it can
disagree without either reporting a problem.

\textbf{The compatibility mechanism and its consequence.} Most hybrid
designs achieve backward compatibility by placing the post-quantum
material where an old verifier may ignore it: a non-critical X.509
extension, or a second certificate bound to the first. RFC 5280
\cite{r1} requires a verifier to reject a certificate carrying a
critical extension it does not recognize, but permits it to ignore a
non-critical one. That permission also lets a verifier validate the
classical path of a hybrid certificate, return accept, and never let the
post-quantum material affect the outcome. The accept is a correct
classical result under RFC 5280, not evidence that the post-quantum
material was checked. By \emph{hybrid authentication} we mean the
stronger judgment a relying party is usually after: that the
post-quantum material was identified, verified, able to change the
verdict had it failed, and valid at the time of validation.

\textbf{A concrete case.} An operator requiring post-quantum protection
deploys hybrid certificates; a client connects, the TLS library
validates the chain, and validation succeeds. In every configuration we
tested that accepted such a certificate on its default path, the library
validated the classical signature and never let the post-quantum extension affect the verdict, returning what it would have returned for a certificate
carrying no post-quantum material, and nothing in that result
distinguishes the two cases. We call this \emph{silent
promotion}: a relying party treats a classical validation result as
though post-quantum authentication had been established. The scenario is
illustrative; what we test is the verifier's behavior, not how an
application reports it.

\textbf{What we found.} Across the schemes that carry the post-quantum
material separably, every stack that parses the certificate validates
the classical path and accepts. We then invalidated the post-quantum
evidence in each separable scheme, leaving the classical evidence
untouched, and put both versions to every configuration: all 27 cells
returned the same verdict, so in none did that evidence bear on the
outcome (Section~\ref{sec:6.1}).
This is not a transitional gap in post-quantum support. Four of the
stacks demonstrably verify post-quantum signatures elsewhere in the same
path, using ML-DSA, the Module-Lattice-Based Digital Signature Algorithm
standardized as FIPS 204 \cite{r4}, and still do not require a hybrid
certificate's post-quantum material to be valid. The gap is structural:
it follows from an extension a verifier may ignore, from hybrid designs
built on that permission, and from the absence of any interface through
which a relying party can require that evidence and learn whether the
requirement was met (Section~\ref{sec:9.1}). That requirement is an operational
policy of the deployment, made explicit as the parameter P0 through P3
in Section~\ref{sec:4.2}, and an operator cannot manage a distinction the verifier
does not surface. The clearest case is a certificate whose bound post-quantum
certificate has been revoked: the classical certificate remains valid,
the revocation lies outside the validated path, and the default
path accepts. All nine configurations accept it on the classical
evidence and none consults the bound certificate, so revoking that
certificate changes no verdict, which we confirmed on one stack with
revocation checking enabled (Section~\ref{sec:7.2}).
The published, retrievable revocation never reaches the verdict, a
certificate lifecycle-management failure.

\textbf{Contributions.} We make four.

\begin{enumerate}
\item A specification-derived model of hybrid certificate validation
(Sections~\ref{sec:2} and \ref{sec:4}), separating four layers: parsing the post-quantum
material, verifying it, letting the verification change the verdict, and
checking its lifecycle status. It states the judgment that distinguishes
a conformant classical acceptance from hybrid authentication.
\item A systematic test across eight stacks and six schemes (Section~\ref{sec:6})
separating capability from enforcement.
\item A lifecycle result for a revoked bound post-quantum certificate
(Section~\ref{sec:7}), with reproducible artifacts.
\item A policy-parametric validation contract (Section~\ref{sec:8}): it takes a
relying party's policy as input, states what a verifier must
establish before an acceptance may be reported as hybrid authentication,
and returns a result labelled with the kind of evidence checked; Section~\ref{sec:9.1} grounds it in the interface argument.
\end{enumerate}
\section{Background}\label{sec:2}
\subsection{Two standards, and what deployed verifiers implement}\label{sec:2.1}
ITU-T Recommendation X.509, published jointly as ISO/IEC 9594-8, defines
the certificate format; the 2019 edition is current. RFC 5280 \cite{r1},
the IETF profile of X.509 for the Internet PKI, fixes which fields and
extensions a conforming implementation must handle and specifies the
path-validation procedure. The alternative-key and alternative-signature
extensions one hybrid design uses were added in ITU-T X.509 \cite{r2}
(2019) clause 9.8. Three of the configurations we test parse and expose
them without acting on them. Two can verify them, one of those three
through a separate API and one through a non-default build, neither on
its default path.

\subsection{Hybrid certificate designs}\label{sec:2.2}
A \emph{hybrid certificate} carries post-quantum key and signature
material alongside classical material, so that verifiers predating
post-quantum algorithms continue to work while those supporting them can
obtain post-quantum assurance. RFC 9794 \cite{r5} calls this arrangement
post-quantum/traditional (PQ/T) hybrid; we write \emph{classical} for
its \emph{traditional} component. We use \emph{post-quantum evidence}
for the post-quantum key and signature material a hybrid design binds to
a certificate, in whatever form the design gives it.

A certificate whose subject key and issuer signature are both
post-quantum, with no classical component, is not a hybrid. We use it
only as a capability baseline: whether a stack can process post-quantum
algorithms at all (Section~\ref{sec:6.3}).

\textbf{Atomic designs} fuse the classical and post-quantum material
into one object that is verified as a unit. The composite signature
scheme \cite{r6} combines several signature algorithms under one
algorithm identifier, and its specification requires every component
signature to be verified. A verifier either implements the composite
algorithm and verifies both components, or does not recognize the
algorithm identifier and fails to process the certificate. There is no
path through which it accepts while disregarding the post-quantum
component.

\textbf{Separable designs} keep the post-quantum evidence in a component
that classical path validation can complete without; we test three:

\begin{itemize}
\tightlist
\item
  \textbf{Catalyst} \cite{r7} places, in extensions of a single
  certificate, an alternative public key and an alternative signature
  computed with a post-quantum algorithm. The certificate's ordinary
  \texttt{signatureValue} remains classical and is what an unmodified
  verifier checks. Clause 9.8 permits these extensions to be marked
  critical or non-critical and recommends non-critical, which our corpus
  follows. Clause 7.2.2 obliges a relying party that has migrated to
  support the alternative algorithms to verify the alternative digital
  signature, and the same clause defines the encoding over which that
  signature is computed (quoted in Section~\ref{sec:9.1}).
\item
  \textbf{Chameleon} \cite{r8} places in one certificate, the base
  certificate, a non-critical extension carrying a descriptor from which
  a second certificate can be reconstructed. The reconstructed
  certificate differs from the base in its key and signature algorithm,
  so a classical base certificate can carry a post-quantum counterpart.
\item
  \textbf{Related certificates} \cite{r3} issue two independent
  certificates for the same subject, one classical and one post-quantum,
  bound by a \texttt{RelatedCertificate} extension carrying a hash of
  the other. Each certificate has its own validity period and revocation
  status. RFC 9763 \cite{r3} describes the endpoint's check, comparing a
  recomputed hash against the extension, but states that how to proceed
  on the outcome is outside its scope and depends on each peer's policy.
\end{itemize}

In all three, a verifier can complete classical path validation without
processing the post-quantum evidence; in the atomic design it cannot.

\subsection{Terminology used in this paper}\label{sec:2.3}
We \emph{validate} certificates and paths, following the usage of RFC
5280, and we \emph{verify} signatures and other individual facts. We say
a verifier \emph{identifies} an extension when it parses and exposes it,
which is what we observe. We reserve \emph{recognized} for RFC 5280's
own undefined term in its obligation to process a recognized
non-critical extension (Section~\ref{sec:4.4}).
\section{Threat Model and Security Goal}\label{sec:3}
Throughout, the relying party we consider has deployed hybrid
certificates for post-quantum protection and therefore requires the
post-quantum evidence to be checked. In the policy scheme of Section~\ref{sec:4.2}
this is P2, which is not a construct of this paper. NIST IR 8547, at
this writing an initial public draft, sets timelines for deprecating the
classical signature algorithms \cite{r22}; ETSI gives migration-planning
guidance \cite{r23}; ANSSI recommends hybrid mechanisms wherever
post-quantum protection is sought during the transition \cite{r24}. A
relying party following such guidance holds the stance P2 names.

\subsection{Two adversaries}\label{sec:3.1}
\textbf{M1} controls which of the certificates already issued to a
server reach the relying party, through an active network position or
control of the server's certificate selection. It cannot forge a
classical signature, obtain a certificate of its choosing from a
certification authority, or alter a certificate without invalidating it,
so every certificate it presents was issued by a real authority and
validates under RFC 5280 \cite{r1}. Its goal is that a relying party
intending hybrid authentication accept on the classical certificate
alone.

\textbf{M2}, the adversary a cryptographically relevant quantum computer
would enable, can forge RSA or ECDSA signatures, including a
certification authority's, but not a post-quantum signature such as
ML-DSA. Unlike M1, it can construct certificates that validate
classically without any authority having issued them. Its goal is to
impersonate a subject by forging the classical component, leaving the
post-quantum evidence as the only protection.

The same verifier behavior defeats a relying party under both: accepting
on classical evidence without requiring the post-quantum evidence to be
present, verified, and outcome-bearing. Our tests instantiate M1, which
applies during the migration period, before a cryptographically relevant
quantum computer exists. M2 is why the gap matters: once such a machine
exists, a verifier that accepts on classical evidence alone provides no
protection at all.

\subsection{Withholding, not tampering}\label{sec:3.2}
The post-quantum evidence in a separable design cannot be deleted from a
certificate. In Catalyst and Chameleon it sits in a non-critical
extension inside the signed portion of the certificate; in Related
certificates the binding hash sits in a signed extension. Removing or
altering either invalidates the issuer's signature, and the certificate
fails to validate.

What an adversary controls is not a certificate's content but which
certificates reach the verifier. An M1 adversary presents a legitimately
issued classical-only certificate for the same subject and withholds the
post-quantum one. The same outcome arises without any adversary when the
hybrid certificate is presented to a verifier whose path never examines
the post-quantum evidence, which is the case our tests instantiate. In
both, every certificate presented validates under RFC 5280 and the
post-quantum protection is not part of the judgment. We therefore speak
of \emph{withholding} the post-quantum evidence rather than of stripping
it, since nothing is removed from a signed structure.

\subsection{What we do and do not claim about conformance}\label{sec:3.3}
We do not describe the behaviors we report as vulnerabilities in the
implementations, because in most cases no implementation departs from a
standard. A verifier that ignores an unrecognized non-critical extension
and accepts is doing what RFC 5280 permits. The failure occurs one level
up, when that conformant classical result is treated as though the
post-quantum evidence had been checked.

This is a statement about implementations, not about the standards.
Scheme specifications state what should be checked, but none attaches a
relying party's hybrid requirement to the called interface, and the gap
we report follows from that absence rather than from any implementation
error (Section~\ref{sec:9.1}). Where an implementation's behavior sits in
tension with a stated requirement we say so and identify it; where the
specifications do not settle the question we mark the case as unsettled
rather than declaring a violation.

\subsection{Security goal}\label{sec:3.4}
A relying party under policy P2 requires that an accepting validation
result be reported as hybrid authentication only if the post-quantum
evidence was identified, verified, outcome-bearing, and current, meaning
valid and not revoked at the same validation time as the classical
evidence. If any of these does not hold, the result may be a correct
classical validation, but must not be reported as a hybrid one.
\section{A Model of Hybrid Certificate Validation}\label{sec:4}
Every judgment traces to a clause of RFC 5280 \cite{r1}, RFC 9763 \cite{r3},
or a scheme specification rather than to the behavior of any toolkit.

\subsection{What classical path validation establishes}\label{sec:4.1}
Path validation (RFC 5280, Section~\ref{sec:6}) builds a certification path from
an end-entity certificate to a trust anchor and evaluates signatures,
validity periods, name constraints, and certificate policies, together
with whatever revocation information the relying party supplies. It
returns accept or reject for the end-entity certificate.

The treatment of extensions is governed by the criticality rule (RFC
5280, Section~\ref{sec:4.2}):

\begin{quote}
A certificate-using system MUST reject the certificate if it encounters
a critical extension it does not recognize or a critical extension that
contains information that it cannot process. A non-critical extension
MAY be ignored if it is not recognized, but MUST be processed if it is
recognized.\footnote{RFC 3280 \cite{r11}, the predecessor profile, stated
  only the permission to ignore an unrecognized non-critical extension;
  RFC 5280 added the obligation to process a recognized one.}
\end{quote}

Path validation therefore establishes the binding recorded in the
certificate's standard fields, the subject public key and the chain of
issuer signatures, and nothing about material carried outside them.

\subsection{The relying party's policy}\label{sec:4.2}
The extension or descriptor carrying the post-quantum evidence is non-critical by design: marking it critical would force
every verifier that does not implement the extension to reject,
defeating the compatibility the designs exist to preserve. RFC 9763
states that the \texttt{RelatedCertificate} extension SHOULD NOT be
marked critical, because marking it critical would severely impact
interoperability. For any verifier that does not recognize the
extension, the post-quantum evidence has therefore been removed from the
accept-or-reject decision as a matter of structure, not of
implementation quality.

It follows that ``this certificate authenticates a
post-quantum-protected identity'' is not a conclusion RFC 5280 path
validation reaches. It is a further judgment layered on classical
acceptance; the obligations that bear on it, in the documents we
examined, are not addressed to an RFC 5280 path validation (Section~\ref{sec:9.1}).

We therefore model the relying party's stance as an explicit parameter
with four settings:

\begin{itemize}
\tightlist
\item
  \textbf{P0 (legacy):} classical path validation only.
\item
  \textbf{P1 (opportunistic):} check the post-quantum evidence when it
  is present, tolerate its absence. This is opportunistic security in
  the sense of RFC 7435 \cite{r26}.
\item
  \textbf{P2 (hybrid-required):} both classical and post-quantum
  evidence must be present and valid.
\item
  \textbf{P3 (continuity):} an identity previously established as
  hybrid may not silently regress to classical-only.
\end{itemize}

Under P0 a verifier that validates the classical path and accepts is
fully conformant with RFC 5280. Under P2 the same acceptance does not
establish hybrid authentication. P2 is a relying-party policy, not a
requirement of any standard, and RFC 9763 is explicit that the choice
lies in policy.

\subsection{Four layers}\label{sec:4.3}
The gap between a classical acceptance and hybrid authentication
decomposes into four questions, each answered differently by default
path validation and by a P2 relying party (Table~\ref{tab:1}).

\textbf{Identification} asks whether the verifier parses the
post-quantum evidence at all. Because RFC 5280 permits ignoring an
unrecognized non-critical extension, a verifier may accept without
registering that it was present. \textbf{Enforcement} asks whether
identified evidence is checked and whether that check can change the
verdict; we call evidence \emph{outcome-bearing} when a failure in it
would turn an otherwise accepting result into a non-accepting one. RFC
5280 obliges a verifier to process a recognized non-critical extension,
and ITU-T X.509 \cite{r2} requires a relying party that has migrated to
verify the alternative signature (clause 7.2.2), but neither states what
a failed verification does to the verdict, so identification does not
entail enforcement. That silence is part of the gap Section~
ef{sec:9.1} states.
\textbf{Policy} asks what a successful
validation reports, a classical result or a hybrid one. It is the
relying party's P0 to P3 choice, a property of the deployment rather
than of the certificate. \textbf{Lifecycle} asks which evidence falls
within the scope of validity and revocation checking; path validation
examines the certificates on the validated path, so when the evidence
lives in a separate certificate, as it does for Related certificates,
that certificate need not be on it and its status is not consulted
(Section~\ref{sec:7}).

\begin{table*}[!t]
\caption{The four layers, contrasting what RFC 5280 path validation establishes by default with what a hybrid-required relying party needs.}
\label{tab:1}
\centering
\footnotesize
\begin{tabular}{@{}
  >{\raggedright\arraybackslash}p{(\textwidth - 6\tabcolsep) * \real{0.2500}}
  >{\raggedright\arraybackslash}p{(\textwidth - 6\tabcolsep) * \real{0.2500}}
  >{\raggedright\arraybackslash}p{(\textwidth - 6\tabcolsep) * \real{0.2500}}
  >{\raggedright\arraybackslash}p{(\textwidth - 6\tabcolsep) * \real{0.2500}}@{}}
\toprule
Layer & Question & Default answer under RFC 5280 & What a P2 relying party requires \\
\midrule
Identification & Is the post-quantum evidence parsed at all? & An
unrecognized non-critical extension may be ignored & The evidence must
be identified \\
Enforcement & Is it checked, and is it outcome-bearing? & Not guaranteed
by path validation & It must be verified, and its failure must prevent
an accepting hybrid result \\
Policy & What does a successful validation report? & A classical
path-validation result & A result labelled with the policy it
satisfies \\
Lifecycle & What is in validity and revocation scope? & Certificates on
the validated path & Every required certificate and binding, at a common
validation time \\
\bottomrule
\end{tabular}
\end{table*}
\subsection{Five outcomes}\label{sec:4.4}
A verifier presented with a hybrid certificate produces one of five
outcomes. We use these labels throughout Sections~\ref{sec:6} and \ref{sec:7}.

\textbf{Hybrid-verified.} The verifier identifies the post-quantum
evidence, verifies it, and lets it bear on acceptance. This is the only
accepting outcome that establishes hybrid authentication under P2.

\textbf{Loud-fail.} The verifier cannot parse or process the structure
and rejects. No acceptance is issued, so no false assurance arises from
this call; the failure is visible to the caller.

\textbf{Classical-accept.} The verifier validates the classical path and
accepts without checking the post-quantum evidence. This is conformant
with RFC 5280, and it does not establish hybrid authentication.

\textbf{Identified-but-not-enforced.} The verifier parses the hybrid
extension and exposes it, but its default path does not let the evidence
affect the outcome. RFC 5280 \cite{r1} obliges a verifier to process an
extension ``recognized by the system'' without defining that term, so a
stack that parses and continues has a defensible reading. ITU-T X.509
\cite{r2}, which defines these extensions, is the closer authority and
states in clause 7.3: ``When a relying party recognizes and is able to
fully process an extension, then the relying party shall process the
extension regardless of the value of the criticality flag.'' All four of
the cells we place in this outcome are at stacks that verify the
alternative algorithm elsewhere on the same path (Section~\ref{sec:6.3}). We still
record these cases as unsettled rather than as violations, because
whether a parser that exposes an extension is thereby able to fully
process it remains a reading rather than a settled matter. We use
\emph{identified} for what we observe (Section~\ref{sec:2.3}).

\textbf{Cross-implementation interoperability failure.} The verifier
does enforce the post-quantum evidence, but rejects a certificate that
another implementation produces and accepts. We report the divergence
without adjudicating which implementation is correct (Section~\ref{sec:6.4}).

These five name what we observe a verifier do. Section~\ref{sec:8} defines a
separate vocabulary for what a contract returns: accept-classical,
accept-hybrid, reject, and indeterminate. The two should not be
confused. An observed \emph{classical-accept} is a verifier accepting on
the classical evidence; a contract result of \emph{accept-classical} is
a statement that such an acceptance is all the policy permits us to
report.

The two pure post-quantum profiles lie outside these five; they serve as
a capability baseline (Section~\ref{sec:6.3}), labelled \emph{PQ-key} and
\emph{PQ-sig} in Table~\ref{tab:4}.

\subsection{The two judgments we apply}\label{sec:4.5}
Each observation is interpreted against two judgments, applied as
described in Section~\ref{sec:5.3}.

The first asks whether RFC 5280 alone permits the observed verdict, with
three possible values: conformant, non-conformant, or unsettled. We
reserve unsettled for the identified-but-not-enforced case described
above.

The second asks what verdict a relying party under a stated policy
should reach. It is produced by a reference procedure of our own rather
than by any tested stack, and that procedure is a policy engine: it
applies the policy parameter, together with a lifecycle rule requiring
every required certificate to be valid and not revoked at a common
validation time, to the recorded evidence state of a case, meaning which
certificates are present, whether each path validates and each required
signature verifies, whether a binding matches, and each certificate's
validity and revocation status. Its rules are grounded in the
specifications and parametric in the policy, and Section~\ref{sec:8} states them
as a contract.

A verifier can be entirely conformant with RFC 5280 and still leave a P2
relying party without the protection it intended.\section{Method}\label{sec:5}
For classical validation the specification and the deployed behaviour
are expected to coincide, and a divergence is a defect. For hybrid
certificates the specification leaves the choice open, so a stack that
ignores the post-quantum evidence and one that enforces it are both
consistent with RFC 5280. Which one a relying party gets has to be
observed rather than derived, which is why we test software.

\subsection{Stacks}\label{sec:5.1}
We tested eight validation stacks drawn from seven independently
developed codebases (Table~\ref{tab:2}). A \emph{validation stack} is a library
together with the specific entry point we exercise; the stacks outnumber
the codebases because oqs-provider adds the Open Quantum Safe provider
to OpenSSL and reuses its path-validation code.

\begin{table*}[!t]
\caption{The validation stacks tested, with versions and build provenance.}
\label{tab:2}
\centering
\footnotesize
\begin{tabular}{@{}
  >{\raggedright\arraybackslash}p{(\textwidth - 6\tabcolsep) * \real{0.2200}}
  >{\raggedright\arraybackslash}p{(\textwidth - 6\tabcolsep) * \real{0.1200}}
  >{\raggedright\arraybackslash}p{(\textwidth - 6\tabcolsep) * \real{0.3600}}
  >{\raggedright\arraybackslash}p{(\textwidth - 6\tabcolsep) * \real{0.3000}}@{}}
\toprule
Stack & Version & Where the build came from & Notes \\
\midrule
OpenSSL & 3.5.7 & upstream source, tag \texttt{openssl-3.5.7} & native
ML-DSA support \\
oqs-provider (Open Quantum Safe) & 0.11.0 & upstream source; liboqs
0.15.0 & over OpenSSL; experimental \\
GnuTLS & 3.7.3 & distribution package \texttt{gnutls-bin
3.7.3-4ubuntu1.9}, Ubuntu 22.04 LTS & \\
Mozilla NSS & 3.98 & distribution package \texttt{libnss3-tools
2:3.98-0ubuntu0.22.04.4}, Ubuntu 22.04 LTS & via \texttt{vfychain},
classic engine \\
Go \texttt{crypto/x509} & 1.26.4 & upstream release tarball & \\
Python \texttt{cryptography} (pyca) & 49.0.0 & PyPI
\texttt{cryptography==49.0.0} & \\
Bouncy Castle (Java) & 1.84 & Maven \texttt{bcprov}, \texttt{bcpkix},
\texttt{bcutil-jdk18on} & alternative-signature verification exposed as a
separate API \\
wolfSSL & 5.9.2 & upstream source, tag \texttt{v5.9.2-stable}, built
twice & alternative-signature enforcement under
\texttt{WOLFSSL\_DUAL\_ALG\_CERTS} \\
\bottomrule
\end{tabular}
\end{table*}
The two frames in the build-provenance column are deliberate, and we state the
consequence rather than leave it to be inferred. Six stacks are pinned
to a chosen upstream release, so they report what a deployment gets when
it tracks the project. GnuTLS and NSS are the packages an Ubuntu 22.04
LTS system installs, so they report what a deployment gets when it
tracks the distribution. Both populations exist, and the second is
large. The cost is that the two rows are older than the other six, and
one result depends on them: the capability count of Section~\ref{sec:6.3} rests in
part on GnuTLS and NSS returning unsupported on the post-quantum
profiles. A current upstream GnuTLS or NSS could move that count
upward. It would not change the enforcement result, which is measured
per configuration and holds for every configuration that accepts,
including the six pinned to current upstream releases.

Every stack is driven through its standard path-validation entry point
with the relying party's trust anchors supplied, and we record the
verdict returned \textbf{without} invoking any additional
alternative-signature or paired-certificate verification the stack may
also expose, because that is what an application obtains unless it
deliberately opts in. Revocation checking is off in the primary matrix,
where every certificate is valid and unrevoked, and supplied only in the
lifecycle runs of Section~\ref{sec:7}.

\begin{table*}[!t]
\caption{The Entry Point Exercised for Each Configuration}
\label{tab:3}
\centering
\footnotesize
\begin{tabular}{@{}
  >{\raggedright\arraybackslash}p{(\textwidth - 6\tabcolsep) * \real{0.1600}}
  >{\raggedright\arraybackslash}p{(\textwidth - 6\tabcolsep) * \real{0.3400}}
  >{\raggedright\arraybackslash}p{(\textwidth - 6\tabcolsep) * \real{0.2400}}
  >{\raggedright\arraybackslash}p{(\textwidth - 6\tabcolsep) * \real{0.2600}}@{}}
\toprule
Configuration & Entry point & Purpose and hostname & Revocation inputs \\
\midrule
OpenSSL & \texttt{openssl verify}, trust anchor and intermediate
supplied & default purpose; no hostname check & accepted
(\texttt{-crl\_check}) \\
oqs-provider & the same call with the provider loaded & default purpose;
no hostname check & accepted \\
GnuTLS & \texttt{certtool} verify against the supplied chain & no
purpose enforced; no hostname check & accepted \\
Mozilla NSS & \texttt{vfychain} over a prepared database, classic
engine & \texttt{certUsageSSLServer}; no hostname check & accepted, with
limits \\
Go \texttt{crypto/x509} & \texttt{Certificate.Verify} with a root pool
and validation time & server authentication; \textbf{hostname checked}
& \textbf{none accepted} \\
Python \texttt{cryptography} & the server verifier built from a policy
builder & server authentication; \textbf{hostname checked} &
\textbf{none accepted} \\
Bouncy Castle & JCA \texttt{CertPathValidator}, PKIX, with the Bouncy
Castle provider & PKIX defaults, no purpose set; no hostname check &
accepted, \emph{explicitly disabled} \\
wolfSSL, default build & \texttt{wolfSSL\_CertManagerVerify} after
loading the anchors & no purpose enforced; no hostname check &
accepted \\
wolfSSL, enforcing build & the same call, compiled with dual-algorithm
support & as above, plus alternative-signature enforcement &
accepted \\
\bottomrule
\end{tabular}

\vspace{3pt}
\begin{minipage}{\textwidth}
\footnotesize
The last two columns record differences we did not normalize away. Two stacks check the hostname because their standard entry point is a server verifier and offers no path-only mode; two accept no revocation input at all; and the Bouncy Castle call disables revocation explicitly.
\end{minipage}
\end{table*}
Two of those differences matter for reading the results. Go and
python-cryptography expose a server verifier rather than a bare path
validator, so their runs carry hostname checking and a server-auth
purpose: their entry point is stricter, not weaker, and they accept the
hybrid certificates anyway. They also accept no revocation input through
this interface, which bounds where the Section~\ref{sec:7} lifecycle result can be
reproduced (Section~\ref{sec:11}).

Two stacks expose enforcement as an option, which we treat as a
controlled variable. Bouncy Castle offers alternative-signature
verification only through a separate API its default path does not call;
wolfSSL can compile it into its standard path, so we report wolfSSL in
two builds, a default build and an enforcing build, both with
post-quantum primitive support, which is what makes the comparison
informative. Counting wolfSSL's two builds separately, this yields 54
standard-entry-point observations over the six profiles. The released
results add four wolfSSL alternative-signature controls and seven
lifecycle observations, for 65 in the primary record, and a contrast set
of 27 valid-and-forged pairs across the three separable schemes and all
nine configurations (Section~\ref{sec:6.1}).

\subsection{The test corpus}\label{sec:5.2}
The corpus covers six certificate profiles, one per scheme: two pure
post-quantum baselines and four hybrid designs, atomic composite and the
three separable designs Catalyst, Chameleon, and Related. We count
profiles throughout, Table~\ref{tab:4} uses the scheme name as each profile's
label, and all are issued under three-tier chains with a common
end-entity profile.

A set of invalidated variants, one per separable scheme, lets us ask
whether a stack enforces the binding rather than merely accepting a
well-formed certificate. Each variant destroys exactly the post-quantum
evidence its scheme's binding depends on: the Catalyst alternative
signature, the Related \texttt{RelatedCertificate} digest, or the
Chameleon delta certificate descriptor's signature value. The classical
evidence is left untouched and valid, and Supplementary Material S6
gives the per-scheme construction. We pair each variant with the valid
certificate it was derived from and drive both members through all nine
configurations. A stack that returns the same verdict for both members
has not let the post-quantum evidence bear on that verdict, because the
only difference between them is that evidence.

The Catalyst and Related variants were confirmed independently, using
an implementation that does check the binding when asked. Bouncy
Castle's opt-in alternative-signature verification returns true for the
valid Catalyst certificate and false for the forged one, and the
\texttt{RelatedCertificate} digest matches the bound certificate for the
valid member and not for the mismatched one. The Chameleon variant is
invalid by construction, a bit flip in the delta certificate
descriptor's signature value. The variants are broken in the way we
intend, so a verdict that does not move is a fact about the verifier
rather than about the corpus.

The second construction is a lifecycle pair for Related certificates: an
independently revocable classical and post-quantum certificate bound by
a \texttt{RelatedCertificate} hash. We place the post-quantum
certificate into revoked, expired, and status-unknown conditions while
the classical certificate remains valid, with both-valid and
single-certificate controls.

Conditions that cannot be exercised against a live responder or
transport, such as an unknown OCSP status or an absent peer, are marked
\emph{modeled} rather than tested, and are evaluated only against the
reference procedure.

Constructing the corpus requires issuing certificates, which the M1
adversary of Section~\ref{sec:3.1} cannot do; it is an instrument for placing a
stack in a defined state, not a claim about attacker capability. The
Related lifecycle states arise with no adversary at all, and the
invalidated variants are states M1 cannot produce and M2 can, used to
test the verifier rather than to model an attack.

\subsection{Classifying each observation}\label{sec:5.3}
Three kinds of statement appear in our results, and we keep them apart
throughout. The first is what a stack returned: a success indication, an
error code, or a report that the structure is unsupported, recorded
verbatim by the adapter. The second is the behavioral outcome we assign
to that run: classical-accept, hybrid-verified, loud-fail,
identified-but-not-enforced, or a cross-implementation interoperability
failure (Section~\ref{sec:4.4}). No stack returns these labels. The third is a
contract
result: accept-classical, accept-hybrid, reject, or indeterminate
(Section~\ref{sec:8.1}). No interface we tested returns one, and we never write
that a stack did.

\textbf{How an outcome is assigned.} For the contrast pairs of Section~\ref{sec:5.2} the assignment is direct. The other labels are analyst-assigned
classifications derived from recorded run properties: the recorded basis
of the acceptance, the extension-handling record, and the capability
baseline, all released per observation. They are not direct telemetry of
every point at which a stack did or did not consult the evidence.

\textbf{The reference procedure.} The procedure that produces the second
judgment of Section~\ref{sec:4.5} applies the policy parameter and the lifecycle
rule of Section~\ref{sec:8.3} to the recorded evidence state. Its executable
wiring covers the Related lifecycle cases of Section~\ref{sec:7}; for the
remaining schemes we apply the same rules to that recorded state and
report the P2 expectation as derived. It verifies no signature of any
scheme: signature validity enters as a recorded input, and what it
computes is the binding digest comparison and each certificate's
validity and revocation status. Where it decodes certificates it uses
Bouncy Castle, also a tested stack, for ASN.1 decoding and for
retrieving revocation and validity fields, while the hash comparison and
every policy decision are our own code; Bouncy Castle's default-path
behavior is recorded as an observation like any other. The procedure
does not ask any implementation what a specification means. The one
place an implementation's own reconstruction of signed bytes is used is
the interoperability comparison of Section~\ref{sec:6.4}.

\subsection{Reproducibility}\label{sec:5.4}
We release the corpus generator, a manifest with per-artifact DER
SHA-256 digests, the stack adapters, the annotated reference procedure,
and the recorded observations with raw logs, produced in a scripted,
containerized environment. Supplementary Material S5 gives the corpus
construction, determinism, and environment detail.
\section{Results: Identification and Enforcement}\label{sec:6}
\subsection{The outcome matrix}\label{sec:6.1}
\begin{table*}[!t]
\caption{Default-Path Outcome by Stack and Scheme}
\label{tab:4}
\centering
\scriptsize
\begin{tabular}{@{}
  >{\raggedright\arraybackslash}p{(\textwidth - 16\tabcolsep) * \real{0.2800}}
  >{\raggedright\arraybackslash}p{(\textwidth - 16\tabcolsep) * \real{0.0900}}
  >{\raggedright\arraybackslash}p{(\textwidth - 16\tabcolsep) * \real{0.0900}}
  >{\raggedright\arraybackslash}p{(\textwidth - 16\tabcolsep) * \real{0.0900}}
  >{\raggedright\arraybackslash}p{(\textwidth - 16\tabcolsep) * \real{0.0900}}
  >{\raggedright\arraybackslash}p{(\textwidth - 16\tabcolsep) * \real{0.0900}}
  >{\raggedright\arraybackslash}p{(\textwidth - 16\tabcolsep) * \real{0.0900}}
  >{\raggedright\arraybackslash}p{(\textwidth - 16\tabcolsep) * \real{0.0900}}
  >{\raggedright\arraybackslash}p{(\textwidth - 16\tabcolsep) * \real{0.0900}}@{}}
\toprule
Scheme & OpenSSL & oqs-prov. & pyca & GnuTLS & Go & BC & NSS & wolfSSL \\
\midrule
pure PQ (ML-DSA subject, classical issuer sig) & PQ-key & PQ-key &
PQ-key & PQ-key & PQ-key & PQ-key & Unsup & PQ-key \\
pure PQ (ML-DSA issuer sig) & PQ-sig & PQ-sig & Unsup & Unsup & Unsup &
PQ-sig & Unsup & PQ-sig \\
composite (atomic) & LF & LF & LF & LF & LF & HV & LF & LF \\
Related & C-Acc & C-Acc & C-Acc & C-Acc & C-Acc & C-Acc & C-Acc &
C-Acc \\
Chameleon & C-Acc & C-Acc & C-Acc & C-Acc & C-Acc & IBNE & Unsup &
C-Acc \\
Catalyst & IBNE & IBNE & C-Acc & C-Acc & C-Acc & IBNE & C-Acc &
C-Acc \\
\bottomrule
\end{tabular}

\vspace{3pt}
\begin{minipage}{\textwidth}
\footnotesize
oqs-prov.\ = oqs-provider; BC = Bouncy Castle. PQ-key = accepts a certificate with an ML-DSA subject key under a classical issuer signature (parsing baseline); PQ-sig = verifies an ML-DSA issuer signature (signature-verification baseline); HV = hybrid-verified; LF = loud-fail (rejected as unrecognized); C-Acc = classical-accept; IBNE = identified but not enforced; Unsup = loud-fail from an unsupported structure. The two pure post-quantum rows are capability baselines and lie outside the five outcomes, which apply to hybrid certificates. wolfSSL is shown in its default build; Section~\ref{sec:6.4} covers its enforcing build.
\end{minipage}
\end{table*}
Across the three separable schemes the pattern is uniform. On its default
path, every stack that parses the certificate returns an accepting
verdict, and in no case does the post-quantum evidence affect that
verdict. The cells differ only in whether the stack noticed the
evidence: classical-accept where it did not,
identified-but-not-enforced where it did.

\textbf{The contrast set establishes this directly.} We ran the
invalidated variant of Section~\ref{sec:5.2} beside its valid counterpart in every
one of the 27 cells formed by the three separable schemes and the nine
configurations. \emph{In all 27 the two verdicts were identical.} No
configuration distinguished a certificate whose post-quantum evidence
was sound from one whose post-quantum evidence had been destroyed. This
is the claim stated as a measurement rather than as an inference, and it
covers every accepting cell in Table~\ref{tab:4}, including the four where the
stack does parse the extension.

Two of the 27 cells deserve their own sentence, because in them the
contrast is uninformative rather than negative. Catalyst at wolfSSL's
enforcing build returns a rejection for both members, but for the reason
of Section~\ref{sec:6.4}: that build rejects the corpus Catalyst certificate
outright over an encoding divergence, before any question of enforcement
arises. Chameleon at NSS returns unsupported for both members, and
Section~\ref{sec:6.2} gives the cause, which lies in the subject key rather than
in the descriptor. Neither cell tests whether the binding is enforced,
because neither reaches the binding. The remaining 25 cells are all
accepting cells, and in all 25 the invalidated variant was accepted on
the same terms as the valid one.

The fields that classify each acceptance remain in the released records,
assigned as Section~\ref{sec:5.3} states. They corroborate the contrast rather
than carry it.

Neither accepting outcome establishes hybrid authentication under P2.
Classical-accept is conformant with RFC 5280 \cite{r1} under P0.
Identified-but-not-enforced is the case we leave unsettled, for the
reason given in Section~\ref{sec:4.4}. In both, the acceptance is recorded as resting on the
classical evidence, and the policy rules return a hybrid result only
after checking evidence that the default path did not consult.

\subsection{Where each outcome appears}\label{sec:6.2}
\textbf{Identified-but-not-enforced} appears for Catalyst at OpenSSL,
oqs-provider, and Bouncy Castle, and for Chameleon at Bouncy Castle.
These stacks parse the hybrid extension and do not let it affect
the verdict. All four cells are in the contrast set, and in
each the invalidated variant was accepted exactly as the valid one was,
so the non-enforcement here is observed and not inferred. The
conformance question matters most in these cells, because a stack that
never parses the extension raises none while one that parses it and
proceeds does.

Unsupported appears for NSS on the Chameleon profile, the single
exception to the statement that every stack parsing a separable
certificate accepted it, and the cause is not the
delta certificate descriptor. The Chameleon base certificate carries an
ML-DSA subject key, and NSS 3.98 does not classify that key type,
returning a key-usage error before the descriptor is reached. The same
error appears on the pure post-quantum profiles for the same reason.
NSS therefore never reaches the Chameleon binding, and this cell reports
nothing about whether NSS would recognize the descriptor. This is a
limitation in NSS's handling of the algorithm rather than the interface
gap this paper is about (Supplementary Material S1).

\subsection{The gap is not missing post-quantum support}\label{sec:6.3}
One objection is that these stacks do not support post-quantum
cryptography yet, and that the behavior will disappear as they mature.
Where this can be tested directly, the data contradict it.

The pure post-quantum rows of Table~\ref{tab:4} isolate capability from
enforcement. The ML-DSA issuer-signature baseline asks whether a stack
can verify a post-quantum signature along a path at all, and four can:
OpenSSL, oqs-provider, Bouncy Castle, and wolfSSL. All four nonetheless
return classical-accept or identified-but-not-enforced on the separable
schemes by default, and for these four missing capability cannot be the
explanation. For the rest the point is structural: primitive support and
hybrid enforcement sit at different layers, so acquiring the first would
not by itself supply the second (Section~\ref{sec:9}).

python-cryptography shows the same thing from the other direction: its
path-validation interface returns unsupported for the ML-DSA
issuer-signature baseline, yet the same interface accepts Catalyst,
Chameleon, and Related certificates as classical-accept. What differs is
not the library's capability but where the post-quantum
material sits: on the validation path as a key it must process, or in a
non-critical extension it may ignore.

The risk in the separable designs is therefore not that a verifier
cannot process post-quantum evidence but that it treats a
hybrid certificate as a classical one.

\subsection{Enforcement that does not interoperate}\label{sec:6.4}
Enforcement is available in one of the stacks we tested: wolfSSL's
enforcing build, compiled with alternative-signature enforcement,
does enforce. On certificates it generates itself, it accepts a valid
alternative signature and rejects a forged one while the base
certificate remains valid. Against our corpus, however, it rejects a
Catalyst certificate generated by Bouncy Castle that Bouncy Castle's own
verification accepts. The two implementations reconstruct the bytes
covered by the alternative signature differently, and each is internally
consistent. On the five profiles other than Catalyst the enforcing build
returns exactly what the default build returns. Catalyst is the only
corpus profile carrying an alternative signature, so it is the only
place where compiling the feature in can change anything.

The divergence is asymmetric: Bouncy Castle accepts certificates from
both sources, while wolfSSL's enforcing build accepts only its own. We
report this as
an observed interoperability failure rather than adjudicating which
reconstruction is correct, and we do not attribute it to a
specific field.

This case is unlike the others in this section. In every cell of Table~\ref{tab:4}
the post-quantum evidence fails to bear on the
verdict; here it bears on the verdict exactly as a hybrid-required
relying party would want, the forged alternative signature being caught
and the certificate rejected. What fails is not enforcement but
interoperability. The byte-level evidence is in Supplementary Material
S2, and Section~\ref{sec:9.2} discusses the incentive this creates for an operator
deciding whether to enable enforcement.
\section{When the Bound Certificate Is No Longer Valid}\label{sec:7}
When the post-quantum certificate bound
to a classical certificate is revoked, expired, or of unknown status,
and the classical certificate remains valid, the default path returns
the same acceptance as before. The reference procedure does not.

\subsection{Why Related certificates are the right setting}\label{sec:7.1}
One bounding statement comes first. RFC 9763 is written for multiple
authentications within a protocol: both certificates are presented, and
each is separately path-validated, so a revoked post-quantum certificate
surfaces in its own validation. On what follows from its check, the RFC
is explicit: ``How to proceed with authentication based on the outcome
of this process is outside the scope of this document. Different
determinations may be made depending on each peer's policy regarding
whether both or at least one authentication needs to succeed.'' The
scenario studied here is the single-presentation case that arises
during migration. Either only the classical certificate is presented,
which the withholding argument of Section~\ref{sec:3.2} already motivates, or the
relying party infers hybrid protection from the satisfied binding
alone. The
lifecycle finding characterizes what default path validation establishes
in that case. It is not a defect in the RFC's own procedure.

\begin{figure*}[!t]
\centering
\includegraphics[width=0.92\textwidth]{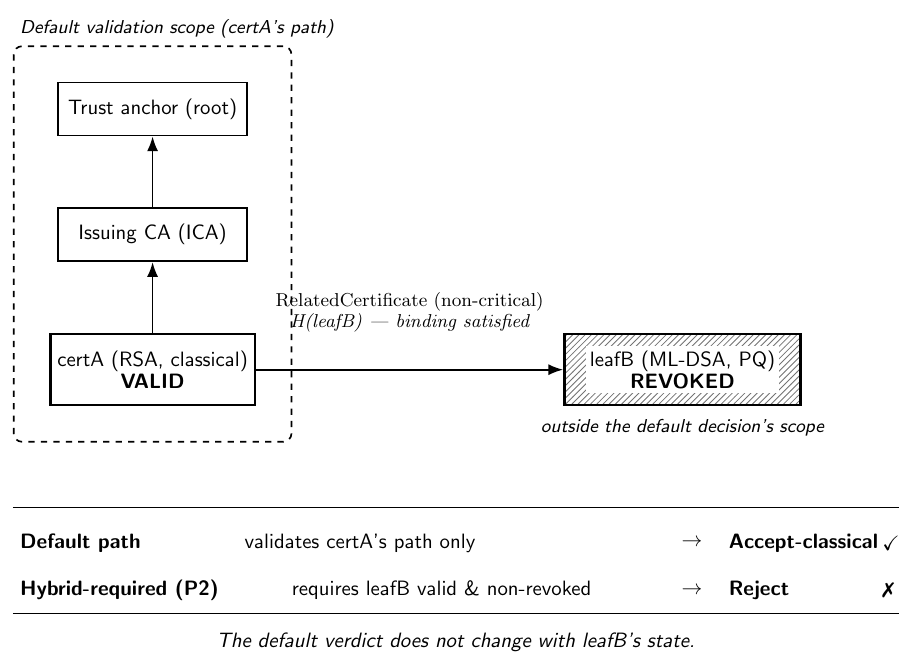}
\caption{A Related certificate pair in which the two certificates are in different lifecycle states. \texttt{certA}, the classical certificate, is valid and its path validates, so the default path accepts. The bound \texttt{leafB} is revoked, but it lies outside the validated path, so its revocation never reaches the verdict. Applying the hybrid-required (P2) rules brings \texttt{leafB} into scope and yields a non-accepting result.}
\label{fig:desync}
\end{figure*}

We use RFC 9763 \cite{r3} Related certificates as the clean case. Two
end-entity certificates are issued for the same subject: a classical
\texttt{certA} using RSA and a post-quantum \texttt{leafB} using ML-DSA,
with \texttt{certA} carrying a non-critical \texttt{RelatedCertificate}
extension binding it to a hash of \texttt{leafB} (Fig.~\ref{fig:desync}).

We include Chameleon as a replication rather than an equal case. Its
paired certificate is a descriptor inside a single issued certificate,
so whether the reconstructed certificate is independently revocable is
arguable; we build no independently revoked Chameleon corpus, and the
lifecycle result rests on Related certificates alone.

In every case below the \texttt{RelatedCertificate} hash matches. The
binding, as a structural assertion, holds. The question is whether a
satisfied binding together with a valid classical certificate amounts to
hybrid authentication when the bound post-quantum certificate is not
itself valid.

\subsection{What happens as \texttt{leafB}'s state varies}\label{sec:7.2}
Table~\ref{tab:5} holds \texttt{certA} valid throughout and varies the state of
\texttt{leafB}. It reports the default-path verdict and the verdict of
the P2 reference procedure. The certificate presented to the verifier is
\texttt{certA} in all five rows, and no configuration consults
\texttt{leafB}, so the acceptance of \texttt{certA} recorded for the
Related profile in all nine configurations (Table~\ref{tab:4}, Section~\ref{sec:6.4})
covers every row rather than each row being driven separately; the
revocation-enabled runs are on OpenSSL. The
markers in the default-path column state what covers each cell. That
invariance is the point of the table: the default verdict does not
change with \texttt{leafB}'s state, which is why a change in that state
goes unnoticed.

\begin{table*}[!t]
\caption{Related-Certificate States with \texttt{certA} Valid in Every Row}
\label{tab:5}
\centering
\footnotesize
\begin{tabular}{@{}
  >{\raggedright\arraybackslash}p{(\textwidth - 6\tabcolsep) * \real{0.2500}}
  >{\raggedright\arraybackslash}p{(\textwidth - 6\tabcolsep) * \real{0.2500}}
  >{\raggedright\arraybackslash}p{(\textwidth - 6\tabcolsep) * \real{0.2500}}
  >{\raggedright\arraybackslash}p{(\textwidth - 6\tabcolsep) * \real{0.2500}}@{}}
\toprule
State of \texttt{leafB} (\texttt{certA} valid) & Default path & P2 reference & Basis for \texttt{leafB} state \\
\midrule
revoked & classical-accept\textsuperscript{a,b} & reject & tested (CRL) \\
expired & classical-accept\textsuperscript{a,b} & reject & tested \\
OCSP status unknown & classical-accept\textsuperscript{b} &
indeterminate & modeled\textsuperscript{c} \\
peer certificate absent & classical-accept\textsuperscript{b} & reject
& modeled\textsuperscript{c} \\
both valid \emph{(control)} & classical-accept\textsuperscript{a} &
accept-hybrid & --- \\
\bottomrule
\end{tabular}

\vspace{3pt}
\begin{minipage}{\textwidth}
\footnotesize
The certificate presented is \texttt{certA} throughout, and no configuration's default path consults \texttt{leafB}. \textsuperscript{a} Directly observed. For the control row this is the Related-profile acceptance recorded in all nine configurations (Table~\ref{tab:4}). For the revoked and expired rows it is the representative OpenSSL run: \texttt{certA} was accepted with revocation checking on and the CRL revoking \texttt{leafB} loaded, and again at a validation time past \texttt{leafB}'s expiry. \textsuperscript{b} Covered by the invariance of that acceptance in the configurations not driven for the row, since \texttt{leafB} is not consulted. The marker \textsuperscript{c} is on the basis column and applies to \texttt{leafB}'s state, not to the default-path cell: that state was modeled rather than driven against real machinery, and only the reference procedure was evaluated on it (Section~\ref{sec:7.2}). Default-path cells use the observational vocabulary of Section~\ref{sec:5.3}; the P2 column uses the contract result types of Section~\ref{sec:8.1}. The reference procedure returns a non-accepting verdict for every state in which the two certificates are out of step, and returns a hybrid acceptance only for the control.
\end{minipage}
\end{table*}
Two things in this table carry the argument. The first is that the
default path accepts in every row, for the reason established in Section~\ref{sec:6}: the validated path is \texttt{certA}'s, \texttt{leafB} is not on it,
and \texttt{leafB}'s condition never enters the decision. The policy
rules, which require every certificate involved to be path-valid and not
revoked at a common validation time, return a non-accepting result in
each of those rows.

The second is the control row. There too the default path accepts on the
classical evidence rather than returning a hybrid result, and its
verdict is identical whether or not \texttt{leafB} is valid, which is
why a later revocation goes unnoticed. The policy rules return a hybrid
result for that row, confirming that they accept when they should rather
than rejecting everything.

We distinguish tested from modeled states. Revoked and expired were
exercised against real revocation machinery (Section~\ref{sec:7.3}). For unknown
OCSP status and absent peer certificate we could not drive a live
responder or a peer-omitting transport, so we mark them modeled and
evaluate them against the reference procedure only. The procedure's
verdicts differ between those two rows for a principled reason. An
absent peer certificate leaves a required certificate missing, so the
procedure rejects. An unknown OCSP status leaves the certificate present
but its revocation state undetermined, so the procedure returns
indeterminate; treating unknown as a rejection would assert a status the
verifier does not have. Neither state permits a hybrid acceptance.

\subsection{This is not revocation failing open}\label{sec:7.3}
Table~\ref{tab:5} invites the reading that we have rediscovered revocation
checking that fails open. Three controls rule that out; the flags,
validation times, and submitted certificates are in Supplementary
Material S3.

First, the revocation machinery works: validated on its own, the revoked
\texttt{leafB} is rejected with a revoked-certificate error and the
expired one with an expired-certificate error. Second, the policy rules
do not simply reject everything, returning a hybrid result for the
both-valid control. Third, the acceptance persists with revocation
checking enabled. \texttt{certA} passes with it on, because
\texttt{leafB} is not on \texttt{certA}'s path and its status is never
requested.

The acceptance is therefore a question of scope, not of failing open.
Revocation soft-fail concerns what a verifier does when it cannot
retrieve a status. Here the status is retrievable, and decisive when the
certificate is checked on its own. The certificate whose status it is
simply never enters the judgment.

\subsection{What this means for a relying party}\label{sec:7.4}
The result has a consequence that requires no quantum computer, and so
falls under adversary M1.

An application reading this result as hybrid authentication
records a post-quantum-protected endpoint, while the post-quantum
certificate meant to provide that protection has been revoked. The
reference procedure, given the same certificates, rejects.

Under adversary M2 the consequence is not merely a mislabelled result.
The classical signature is forgeable, so a valid and outcome-bearing
post-quantum certificate is the only thing that would prevent
impersonation. If the bound certificate is revoked, or is absent from
the judgment, a P2 relying party must not accept. The default path
accepts on the classical certificate alone.

In both cases the failure is the one named in Section~\ref{sec:3}: under a
hybrid-required policy, a valid classical result is treated as a hybrid
one that the evidence does not support. Section~\ref{sec:8} states, as a contract
whose policy rules our reference procedure implements for the lifecycle
cases, the discipline that would prevent it.
\section{A Policy-Parametric Validation Contract}\label{sec:8}
What Sections~\ref{sec:6} and \ref{sec:7} show to be missing is not algorithm support. It is
a validation result that tells the relying party which kind of evidence
the acceptance rests on, and that lets the post-quantum evidence change
the verdict when it fails.

This section states that missing piece as a \emph{policy-parametric
validation contract}: a specification of what a verifier must establish,
given a scheme and a relying party's policy, before a validation may be
reported as hybrid authentication. The reference procedure of Section~\ref{sec:5.3} implements its policy rules, which we ran over the lifecycle cases
of Section~\ref{sec:7}. Where the two differ we say so: the contract states what a
conforming verifier would have to check for each scheme, which is more
than our procedure implements, and it is in any case a reference
statement rather than a production verifier (Section~\ref{sec:11}).

\subsection{Policies and result types}\label{sec:8.1}
The contract takes the policy of Section~\ref{sec:4.2} as a parameter and returns
one of four result types: accept-classical, accept-hybrid, reject, and
indeterminate. It does not return a single accept-or-reject bit, because
the failure this paper documents is not a wrong bit but a result of one
kind being read as a result of another: a P0 accept-classical taken for
a P2 accept-hybrid. Table~\ref{tab:6} gives the mapping.

\begin{table}[!t]
\caption{The Contract's Result as a Function of Policy and Evidence State}
\label{tab:6}
\centering
\small
\begin{tabular}{@{}
  >{\raggedright\arraybackslash}p{(\columnwidth - 4\tabcolsep) * \real{0.3333}}
  >{\raggedright\arraybackslash}p{(\columnwidth - 4\tabcolsep) * \real{0.3333}}
  >{\raggedright\arraybackslash}p{(\columnwidth - 4\tabcolsep) * \real{0.3333}}@{}}
\toprule
Policy & Required PQ evidence valid and outcome-bearing & Required PQ evidence invalid, absent, or unsupported \\
\midrule
P0 (classical) & accept-classical & accept-classical \\
P1 (opportunistic) & accept-hybrid & accept-classical permitted;
never accept-hybrid \\
P2 (hybrid required) & accept-hybrid & non-accepting: reject, or
indeterminate if status is unknown \\
P3 (hybrid + continuity) & accept-hybrid & non-accepting; and a later
accept-classical for an identity recorded as hybrid-required is itself
non-accepting \\
\bottomrule
\end{tabular}

\vspace{3pt}
\begin{minipage}{\columnwidth}
\footnotesize
A valid classical path is assumed throughout. P3 presupposes an external continuity state keyed by the relying party's identity notion; we model only the verifier-side implication, not how it is stored or populated (Section~\ref{sec:10}).
\end{minipage}
\end{table}
Reject and indeterminate are both non-accepting for hybrid purposes.
Reject is returned when a required certificate is invalid or absent:
revoked, expired, or not presented. Indeterminate is returned when a
required certificate is present but its status cannot be determined, as
with an unknown OCSP response, so that the contract declines to assert a
revocation status the verifier does not have. Whether to collapse
indeterminate into reject is left to the relying party; neither may be
reported as accept-hybrid.

The contract has a second parameter, the \emph{path scope} at which
post-quantum evidence is required: end-entity only, the issuing path, or
the full path. Our tests instantiate the end-entity scope. Under M2 the
choice matters: if a classical issuer signature is forgeable,
post-quantum evidence at the leaf alone does not establish hybrid
authentication for the whole path. A deployment that needs path-wide
assurance would set a stricter scope. Composing the requirement across a
chain is a dimension of the contract we do not develop here (Section~\ref{sec:11}).

What each policy is worth depends on the adversary of Section~\ref{sec:3.1}, and
Supplementary Material S7 collects the comparison. Under M1, P1 provides
no more protection than P0: the adversary withholds the post-quantum
evidence, P1 tolerates the absence, and the result is accept-classical,
which the policy permits. P1's value is therefore confined to reporting,
since it never labels a result accept-hybrid unless the evidence held.
P2 is the first policy under which withholding no longer yields an
acceptance the relying party can misread.

P3 detects regression and requires continuity state. Its operational
risk depends on where that state comes from, and the two sources differ
in kind; Supplementary Material S7 sets them out. Table~\ref{tab:6} and
Section~\ref{sec:8.4} use the inventory form.

\subsection{What each scheme requires}\label{sec:8.2}
Before reporting accept-hybrid under P1, P2, or P3, the contract
requires the evidence in Table~\ref{tab:7} to be verified and outcome-bearing in
addition to a valid classical path.

\begin{table*}[!t]
\caption{Evidence the Contract Requires Before a Hybrid Result}
\label{tab:7}
\centering
\small
\begin{tabular}{@{}
  >{\raggedright\arraybackslash}p{(\textwidth - 4\tabcolsep) * \real{0.1800}}
  >{\raggedright\arraybackslash}p{(\textwidth - 4\tabcolsep) * \real{0.6500}}
  >{\raggedright\arraybackslash}p{(\textwidth - 4\tabcolsep) * \real{0.1700}}@{}}
\toprule
Scheme & Evidence required for a hybrid result, beyond a valid classical path & Status \\
\midrule
Atomic composite & Both component signatures verify together under the
composite algorithm identifier; failure of either causes the single
verification to fail. & derived \\
Catalyst (X.509 alternative) & The alternative signature verifies under
the issuer's bound alternative public key and alternative signature
algorithm, over the scheme-defined pre-signature encoding; the base
certificate path is valid; and all alternative-signature inputs the
scheme requires are present and bound to the issuing certificate. &
derived \\
Chameleon (delta certificate) & The delta certificate is reconstructed
from the descriptor; the verifier validates, under RFC 5280, whichever
of the base and delta certificates the policy requires; and the
descriptor-derived binding is verified. & stated \\
Related (RFC 9763 \cite{r3}) & Both the classical and the post-quantum
certificate are path-valid, the \texttt{RelatedCertificate} hash binds
them, and both satisfy the lifecycle rule of Section~\ref{sec:8.3};
which related certificates are required is supplied by policy, and a
hash match establishes relatedness, not authorization. & exercised \\
\bottomrule
\end{tabular}

\vspace{3pt}
\begin{minipage}{\textwidth}
\footnotesize
The status records how far we took each row. \emph{Exercised}: the
reference procedure ran the row's policy rules over recorded cases
(Sections~\ref{sec:5.3} and \ref{sec:7}). \emph{Derived}: the
requirement was applied to the recorded evidence state without running
the procedure (Section~\ref{sec:5.3}). \emph{Stated}: the requirement
was not exercised. Our reference procedure does not implement the
Chameleon reconstruction, which is why that row is stated rather than
exercised; whether a reconstructed delta certificate is independently
revocable remains a caveat (Section~\ref{sec:7.1}), carried into the
lifecycle rule.
\end{minipage}
\end{table*}
Atomic composite is the one scheme where a scheme-aware default path
already makes the evidence outcome-bearing, because the evidence lives
inside the single signature the verifier checks; the separable schemes
require the verifier to reach outside the classical path, to evidence it
is otherwise free to ignore. For Catalyst the pre-signature encoding
named in the table is the reconstruction over which the two enforcing
implementations disagreed (Section~\ref{sec:6.4}): the contract requires the
alternative signature to verify over the scheme-defined encoding,
without adjudicating what that encoding is.

\subsection{The lifecycle rule}\label{sec:8.3}
The contract requires every certificate a scheme designates as required
to be within its validity window, on a valid certification path, and not
revoked, all at a single common validation time. Section~\ref{sec:7} showed that
default paths do not apply this rule to the separable schemes: where the
default path returns a classical-accept, the contract returns reject
when a required certificate is revoked, expired, or absent, and
indeterminate when its status cannot be determined. The rule does not depend on which
certificate carries the binding; the orientation control is in
Supplementary Material S4.

The contract adds no cryptography to the schemes. It adds the validation
result that the schemes and RFC 5280 together leave unspecified: a
result labelled with the policy it satisfies, resting on evidence whose
failure would have changed it. Supplementary Material S8 works the
binding through for three of the tested stacks and states what each
retrofit costs.

\subsection{What a management plane does with a labelled result}\label{sec:8.4}
None of this is implemented, and we claim no evaluation of it. The
obstacle to building the three functions below is that no validation
interface we tested returns the two facts they consume.

\textbf{Enrolment and inventory.} A deployment that requires hybrid
authentication for an identity has to record that requirement somewhere.
What is missing is not the record. It is the return path: none of the
interfaces we tested reports whether validation honored it.

\textbf{Audit.} The labelled result supplies that return path: a
validation that returns accept-classical for an identity the inventory
records as hybrid-required is the P3 condition of Section~\ref{sec:8.1}, and an
auditor can query for it from the record alone, without re-running the
validation. Under the current interface that situation leaves no trace,
its verdict being indistinguishable from a hybrid acceptance.

\textbf{Monitoring.} In Section~\ref{sec:7}'s revocation case the verdict does not
change, so nothing a monitoring system watches changes. With a labelled
result the same event produces a visible transition, from accept-hybrid
to a non-accepting result under P2, of the kind operators already act on
for expiry and revocation. The same records
carry a migration measure no verdict we recorded could supply: the share of
validations across an estate that return accept-hybrid rather than
accept-classical, per identity and over time.
\section{Discussion}\label{sec:9}
One reading would set Sections~\ref{sec:6} and \ref{sec:7} aside: libraries
have not yet caught up with post-quantum cryptography, and the behavior
will disappear as they do. But capability and enforcement are different
layers, and maturation operates on the lower one.

\subsection{The obligations exist; nothing connects them to the deployed interface}\label{sec:9.1}
\textbf{The obligations are real, and they differ by scheme.} This is a
gap the standards leave open rather than an error made by implementers
(Section~\ref{sec:3.3}). The composite specification requires every component
signature to be verified. Because the composite algorithm identifies the
certificate's signature, an RFC 5280 \cite{r1} path validation that
checks that signature necessarily invokes that obligation.

For Catalyst, ITU-T X.509 \cite{r2} clause 7.2.2 states the obligation
directly: ``A relying party that has migrated to support alternative
cryptographic algorithms and alternative digital signatures shall verify
the alternative digital signature.'' The same clause defines the
encoding over which that signature is computed, by exclusion of the
\texttt{signature} component and the \texttt{altSignatureValue}
extension from the re-encoded certificate; clauses 9.8.2 through 9.8.4
define the carrying extensions. For Related certificates, RFC 9763
\cite{r3} describes the check an endpoint performs, then states that how
to proceed on its outcome is outside the document's scope and depends on
each peer's policy, and that the mechanism does not by itself effect any
security function.

\textbf{What is missing is an interface that carries the requirement.}
Each obligation presumes a verifier that has already decided to apply
it. None of the specifications we examined --- RFC 5280, ITU-T X.509,
RFC 9763, and the composite and Chameleon drafts --- defines a
path-validation entry point that takes a relying party's hybrid
requirement as input, selects the checks that requirement implies, and
returns a result saying which kind of acceptance was reached.

\textbf{The nearest existing mechanism is certificate-policies
processing.} That claim needs one qualification, because RFC 5280 path
validation does accept relying-party policy inputs. Its procedure takes
a \texttt{user-initial-policy-set} and an
\texttt{initial-explicit-policy} indicator, and it returns a
\texttt{valid\_policy\_tree} (RFC 5280, Section~\ref{sec:6.1}.1). A certification
authority could assert a hybrid-issuance policy identifier in the
certificates it issues, and a relying party demanding that identifier
through \texttt{initial-explicit-policy} would reject certificates that
do not assert it. This machinery is the closest thing the deployed
interface offers to a policy parameter, and it deserves the comparison.
What it matches, however, is CA-asserted identifiers, not evidence. A
certificate can assert the identifier while its alternative signature
is invalid, and policy processing will not notice, because nothing in
it verifies the post-quantum evidence or makes that evidence
outcome-bearing. It does not bring a bound certificate's lifecycle into
the scope of the decision, so the revoked certificate of Section~\ref{sec:7}
stays outside it. And it reports which policies a path is valid under,
not which checks were performed, so a classical acceptance and a hybrid
one still come back indistinguishable. The gap this paper names is what
remains after certificate-policies processing is given full credit.

The ITU-T obligation is conditioned on a relying party that has migrated
to support the alternative algorithms. No interface we tested lets a
deployment declare that state or reports whether a validation acted on
it, so an application cannot establish from a result whether the
obligation was in force. We treat this as a reason the conformance
question is unsettled for the affected cells (Section~\ref{sec:4.4}), not as an
exemption.

\subsection{Why enforcement is not a drop-in fix}\label{sec:9.2}
We do not conclude from the Catalyst divergence of Section~\ref{sec:6.4} that the
specification is ambiguous. We conclude that the encoding has not been
settled by a normative profile or a required conformance suite. The
IETF Hackathon PQC certificates project \cite{r25} maintains a
cross-vendor artifact repository and compatibility matrix whose
artifact-naming scheme provides for Catalyst-format hybrids. That such
a venue exists makes the divergence we observed more significant, not
less: participation is voluntary, the matrix binds no implementation,
and the divergence persists between shipping builds.

Enforcement creates an operational incentive toward the default: an
operator who enables it and finds another vendor's certificates rejected
has a working configuration one build flag away, and it is the
non-enforcing one.

The gap takes two forms:
without enforcement, a hybrid certificate is accepted as a classical one
and nothing is visible; with enforcement, it may not validate across
implementations. ``Just enforce it'' is not a remedy a relying party can
apply unilaterally.

\subsection{The protocol layer does not close the gap}\label{sec:9.3}
A TLS implementer can object that the policy channel already exists one
layer up: TLS 1.3's \texttt{signature\_algorithms\_cert} extension
\cite{r28} lets a client state which signature algorithms it will accept
in the certificate chain.

Two things survive the objection. First, negotiation selects which chain
is presented; it does not bring a bound certificate's lifecycle into
scope. The verifier still validates the presented certificate's own
path, and the revoked bound certificate of Section~\ref{sec:7} lies outside it.
Second, negotiation reports nothing: the handshake succeeds or
fails, and no validation result distinguishes kinds of acceptance. A
client that constrained the algorithms and completed the handshake still
cannot tell, from any result it holds, what kind of acceptance its
verifier reached.

Whole PKI domains have no negotiation layer at all: code signing,
document signing, S/MIME, long-lived device and IoT provisioning, and
offline verification present a certificate without a wire on which to
state a requirement. There the certificate layer is the only policy
channel, and the contract addresses those domains. For
TLS, negotiation governs what is presented; the contract governs what an
acceptance establishes and reports.

\subsection{What would have to change}\label{sec:9.4}
Closing the gap requires action at three layers, and no one of them
suffices.

\textbf{Standards} would have to make the evidence encodings unambiguous
and interoperably testable, through profiles or conformance vectors, so
that enforcing implementations agree (Section~\ref{sec:9.2}); define what
processing a hybrid extension entails, so that the RFC 5280 obligation
has a determinate meaning for evidence of this kind; and state the
lifecycle requirement where a paired certificate is independently
revocable.

\textbf{Libraries} would have to return a result that says which kind of
acceptance was reached, rather than a bare verdict the caller must
interpret. An opt-in enforcement API is a partial answer: an interface
an application must know to call does not protect an application that
does not know to call it. Supplementary Material S8 sketches the binding
for three of the stacks tested.

\textbf{Relying-party interfaces} would have to let a deployment state
that it requires hybrid authentication and receive an answer under that
policy. Today the requirement exists only as intention, and no verifier
we tested can read it or report against it.
\section{Related Work}\label{sec:10}
\textbf{Foundations of hybrid certificates.} Bindel et al.~\cite{r19}
embedded second signatures in non-critical X.509 extensions and
confirmed that the implementations deployed at the time accept such
certificates on the classical component; Kampanakis et al.~\cite{r20}
examined the viability of post-quantum signatures in X.509 certificates
and the protocols that use them, and
Crockett et al.~\cite{r21} prototyped post-quantum and hybrid
authentication in TLS and SSH. That line established the compatibility
half of the separable designs: old verifiers tolerate the added
material. We test whether verifiers that support the post-quantum
algorithms enforce that material by default, show that a bound
certificate's lifecycle falls outside the default decision (Section~\ref{sec:7}),
and state the policy-parametric contract that would carry a relying
party's requirement into the result (Section~\ref{sec:8}). None of these questions
arises while acceptance by legacy verifiers is the property under study.

\textbf{Testing validation, and validators with explicit policies.}
Differential testing, from Frankencerts \cite{r12} to x509-limbo
\cite{r13} and recent post-quantum linting \cite{r14}, generates or
mutates certificates to expose disagreement between validators; Verdict
\cite{r15} states validator correctness against a user-supplied policy.
For a hybrid certificate
there is no single correct verdict, because the intended result depends
on the relying party's policy (Sections~\ref{sec:4.5} and \ref{sec:5.3}), so we interpret
each observation against two judgments rather than one expected result
and do not treat our reference procedure as ground truth. Other work
generates hybrid and composite certificates \cite{r16} and compares the
hybrid X.509 schemes \cite{r17}; our subject is what a relying party has
established when its verifier accepts one.

\textbf{Concurrent work on binding enforcement.} Independently and at
the same time, Lee et al.~\cite{r18} add a binding-enforcement axis to a
transition-cost model to re-rank hybrid strategies. In their
forge-and-strip setting they report Bouncy Castle splitting between its
default path and its opt-in verification API on a forged Catalyst
certificate. Our contrast set reaches that cell independently and finds
the same split: Bouncy Castle's default path accepts our forged Catalyst
certificate on the same terms as the valid one, while the opt-in
alternative-signature call rejects the forged member and accepts the
valid one (Sections~\ref{sec:5.2} and \ref{sec:6.1}).

The wolfSSL comparison needs a distinction the two settings do not share.
They report wolfSSL as the one stack that checks a present alternative
signature by default while being unable to require one, so that a
stripped certificate is accepted. In our measurements that checking is a
compile-time option: the default build does not parse the alternative
signature at all, and only the build compiled with dual-algorithm
support checks it and rejects a forged one (Sections~\ref{sec:6.1} and \ref{sec:6.4}). The
two accounts are compatible if their wolfSSL carried that option, and
the difference matters because what a relying party gets depends on how
its library was built and not on the library alone. We did not test a
stripped certificate. Our variants invalidate evidence that is present
rather than removing it, so the inability to require an absent signature
is their finding and not ours.

Their enforcement measure is a binary axis in a cost model, and they
name the distinction between verifier behavior and relying-party policy
as future work. That distinction is what we supply, together with the
model and contract of Sections~\ref{sec:4} and \ref{sec:8} and the lifecycle result of
Section~\ref{sec:7}, which does not arise in their forge-and-strip setting. They
ask which strategy an operator should choose. We ask what a verifier's
acceptance establishes.

\textbf{Downgrade defenses above this layer.} Two Internet-Drafts let a
server commit to presenting post-quantum certificates and a client
detect a later regression to classical-only \cite{r9,r10}. They
presuppose a relying party that can enforce post-quantum evidence once
it is present, which our results show default verifiers usually cannot
(Sections~\ref{sec:6} and \ref{sec:7}), so the contract of Section~\ref{sec:8} is a precondition for
these defenses rather than an alternative to them.
\section{Limitations}\label{sec:11}
\textbf{Modeled states.} Unknown OCSP status and an absent peer
certificate cannot be driven against real infrastructure; both are
marked modeled and evaluated against the reference procedure only
(Section~\ref{sec:7.2}). Revoked and expired were exercised against real
revocation machinery (Section~\ref{sec:7.3}).

\textbf{Sample coverage.} The eight stacks over seven codebases are a
representative sample, not a census (Section~\ref{sec:5.1}), and a stack outside
it could enforce a hybrid binding by default; Section~\ref{sec:9.1}'s
specification-based argument does not depend on exhaustiveness.

\textbf{Build populations.} GnuTLS and NSS are older Ubuntu 22.04 LTS
builds, the other six pinned upstream releases (Table~\ref{tab:2}). Only the
capability count of Section~\ref{sec:6.3} is sensitive: current upstream builds
could raise it. The enforcement result holds in every accepting
configuration.

\textbf{Lifecycle reproducibility.} Go's \texttt{crypto/x509} and
python-cryptography accept no revocation input through the entry point
an application uses (Table~\ref{tab:3}), so Section~\ref{sec:7}'s lifecycle question cannot
be put to them there and was run on a stack whose entry point accepts
it.

\textbf{Not a production verifier.} The P2 reference procedure, and the
Section~\ref{sec:8} contract it states, are built for traceability to
specification clauses, not performance or feature coverage (Section~\ref{sec:5.3}). Our results instantiate P2, one policy among several, and are
stated conditionally.

\textbf{Specification maturity.} Related certificates are specified in
RFC 9763 \cite{r3}, composite and alternative-signature mechanisms are
standards-track or established, and Chameleon is an expired individual
draft (Section~\ref{sec:7.1}). Changes could require revising Section~\ref{sec:8.2}'s
per-scheme requirements; the gap follows from a pattern the separable
designs share.

\textbf{Interoperability.} In the Section~\ref{sec:6.4} case we establish that the
two implementations reconstruct different pre-signature bytes, each
internally consistent (Supplementary Material S2); we neither locate the
divergence to a field nor determine which reconstruction is correct.

\textbf{Path scope.} Our tests instantiate the end-entity path scope of
Section~\ref{sec:8.1}; hybrid evidence across a whole chain, hybrid intermediates,
mixed chains, and the M2 scope for a forgeable classical issuer
signature were not tested, though the contract admits them as a stricter
path scope.

\textbf{Certificate evidence.} We report what certificate path
validation establishes, not what a completed handshake or application
concludes: we implement no TLS transcript binding, certificate-selection
logic, or application-level fallback, and do not address
negotiation-layer downgrade (Section~\ref{sec:10}).

\textbf{Residual confounds.} We removed certificate-profile confounds
during corpus construction and recorded the superseded observations
(Section~\ref{sec:5.2}; Supplementary Material S1); undetected confounds cannot be
excluded, and the released package and raw logs allow rechecking.

\textbf{Ethics.} The reported behavior is in most cases conformant, so
the appropriate venue is standards and maintainer feedback, not
embargoed disclosure. We introduce no attack technique or exploit
tooling and designate no implementation a violator; we decline to
adjudicate the Section~\ref{sec:6.4} divergence, which no profile or conformance
test settles. The corpus is synthetic, with no human subjects or
personal data.
\section{Conclusion}\label{sec:12}
We asked whether a default path validation on a hybrid X.509
certificate can be read as hybrid authentication. For the separable
designs, Catalyst, Chameleon, and Related, across eight validation
stacks over seven independent codebases, it cannot.

On its default path, every stack that parses the certificate validates
the classical path and accepts. In all 27 cells, three separable
schemes across nine configurations, the invalidated certificate and its
valid counterpart returned the same verdict (Section~\ref{sec:6.1}). The one
configuration compiled to enforce does enforce (Section~\ref{sec:6.4}). All nine
configurations accept a Related certificate on the classical evidence,
so revoking the bound post-quantum certificate does not change the
verdict: its state is never read (Section~\ref{sec:7}).

ITU-T X.509 \cite{r2} and RFC 9763 \cite{r3} state what a verifier
should check, and two stacks implement those checks, but neither on its
default path; no document we examined defines the interface between them, one that
carries a relying party's hybrid requirement into path validation and
reports against it (Section~\ref{sec:9.1}).

Closing the gap requires all three layers to move: standards making the
encodings and enforcement semantics interoperable, libraries returning
a result labelled with the policy it satisfies, and relying-party
interfaces that let a deployment state its hybrid requirement (Section~\ref{sec:9.4}). Until they do, what a default verifier returns will remain a
correct classical result that an application is free to read as a
hybrid one.

%% file: manuscript-ieee.bbl
\begin{thebibliography}{99}
\bibitem{r1} D. Cooper, S. Santesson, S. Farrell, S. Boeyen, R. Housley, W. Polk, Internet X.509 Public Key Infrastructure Certificate and Certificate Revocation List (CRL) Profile, RFC 5280, IETF, 2008. \url{https://doi.org/10.17487/RFC5280}
\bibitem{r2} ITU-T, Recommendation X.509 (10/2019) | ISO/IEC 9594-8: Information technology — Open Systems Interconnection — The Directory: Public-key and attribute certificate frameworks, ITU-T, 2019 (alternative-extension definitions in clauses 9.8.2--9.8.4; verification obligation and encoding in clause 7.2.2).
\bibitem{r3} A. Becker, R. Guthrie, M. Jenkins, Related Certificates for Use in Multiple Authentications within a Protocol, RFC 9763, IETF, 2025. \url{https://doi.org/10.17487/RFC9763}
\bibitem{r4} National Institute of Standards and Technology, Module-Lattice-Based Digital Signature Standard, FIPS 204, NIST, 2024. \url{https://doi.org/10.6028/NIST.FIPS.204}
\bibitem{r5} F. Driscoll, M. Parsons, B. Hale, Terminology for Post-Quantum Traditional Hybrid Schemes, RFC 9794, IETF, 2025. \url{https://doi.org/10.17487/RFC9794}
\bibitem{r6} M. Ounsworth, J. Gray, M. Pala, J. Klaußner, S. Fluhrer, Composite Module-Lattice-Based Digital Signature Algorithm (ML-DSA) for use in X.509 Public Key Infrastructure, Internet-Draft draft-ietf-lamps-pq-composite-sigs-19, IETF, 21 April 2026 (work in progress).
\bibitem{r7} A. Truskovsky, D. Van Geest, S. Fluhrer, P. Kampanakis, M. Ounsworth, S. Mister, Multiple Public-Key Algorithm X.509 Certificates, Internet-Draft draft-truskovsky-lamps-pq-hybrid-x509, IETF, 2018 (expired).
\bibitem{r8} C. Bonnell, J. Gray, D. Hook, T. Okubo, M. Ounsworth, A Mechanism for Encoding Differences in Paired Certificates, Internet-Draft draft-bonnell-lamps-chameleon-certs-07, IETF, 18 October 2025 (expired 21 April 2026).
\bibitem{r9} T. Reddy, J. Gray, Y. Sheffer, X.509 Extensions for PQC or Composite Certificate Hosting Continuity, Internet-Draft draft-reddy-lamps-x509-pq-commit-01, IETF, 25 February 2026 (work in progress).
\bibitem{r10} Y. Sheffer, T. Reddy, PQC Continuity: Downgrade Protection for TLS Servers Migrating to PQC, Internet-Draft draft-sheffer-tls-pqc-continuity-02, IETF, June 2026 (work in progress).
\bibitem{r11} R. Housley, W. Polk, W. Ford, D. Solo, Internet X.509 Public Key Infrastructure Certificate and Certificate Revocation List (CRL) Profile, RFC 3280, IETF, 2002. \url{https://doi.org/10.17487/RFC3280}
\bibitem{r12} C. Brubaker, S. Jana, B. Ray, S. Khurshid, V. Shmatikov, Using Frankencerts for Automated Adversarial Testing of Certificate Validation in SSL/TLS Implementations, in: IEEE Symposium on Security and Privacy (S\&P), 2014, pp. 114–129. \url{https://doi.org/10.1109/SP.2014.15}
\bibitem{r13} C2SP, x509-limbo: A test-vector suite for X.509 path validation [software], n.d. \url{https://github.com/C2SP/x509-limbo} (accessed 30 June 2026).
\bibitem{r14} J.L. Delgado Jiménez, From Public-Key Linting to Operational Post-Quantum X.509 Assurance for ML-KEM and ML-DSA: Registry-Driven Policy, Mutation-Based Evaluation, and Import Validation, arXiv preprint arXiv:2604.17003, 2026.
\bibitem{r15} Z. Lin, M. McLoughlin, P. Singh, R. Brennan-Jones, P. Hitchcox, J. Gancher, B. Parno, Towards Practical, End-to-End Formally Verified X.509 Certificate Validators with Verdict, in: 34th USENIX Security Symposium (USENIX Security '25), 2025, pp. 5035–5052.
\bibitem{r16} N. Ricchizzi, C. Schwinne, J. Pelzl, Applied Post Quantum Cryptography: A Practical Approach for Generating Certificates in Industrial Environments, arXiv preprint arXiv:2505.04333, 2025.
\bibitem{r17} A.C.H. Chen, A Comparative Study of Hybrid Post-Quantum Cryptographic X.509 Certificate Schemes, arXiv preprint arXiv:2511.00111, 2025.
\bibitem{r18} M. Lee, M. Sim, S. Eum, S. Cho, Y. Hyoung, H. Seo, Beyond Size: Do Hybrid PQC Certificates Actually Enforce the Classical–PQC Binding? A Cost-and-Security Study, Cryptology ePrint Archive, Paper 2026/1416, 2026.
\bibitem{r19} N. Bindel, U. Herath, M. McKague, D. Stebila, Transitioning to a Quantum-Resistant Public Key Infrastructure, in: Post-Quantum Cryptography (PQCrypto 2017), LNCS 10346, Springer, 2017, pp. 384–405. \url{https://doi.org/10.1007/978-3-319-59879-6_22}
\bibitem{r20} P. Kampanakis, P. Panburana, E. Daw, D. Van Geest, The Viability of Post-quantum X.509 Certificates, Cryptology ePrint Archive, Paper 2018/063, 2018. \url{https://eprint.iacr.org/2018/063}
\bibitem{r21} E. Crockett, C. Paquin, D. Stebila, Prototyping post-quantum and hybrid key exchange and authentication in TLS and SSH, Cryptology ePrint Archive, Paper 2019/858, 2019. \url{https://eprint.iacr.org/2019/858}
\bibitem{r22} National Institute of Standards and Technology, Transition to Post-Quantum Cryptography Standards, NIST IR 8547 (initial public draft), NIST, November 2024. \url{https://doi.org/10.6028/NIST.IR.8547.ipd}
\bibitem{r23} ETSI, CYBER; Migration strategies and recommendations to Quantum Safe schemes, ETSI TR 103 619 V1.1.1, ETSI, July 2020.
\bibitem{r24} Agence nationale de la sécurité des systèmes d'information, ANSSI views on the Post-Quantum Cryptography transition, position paper, ANSSI, 2022. \url{https://cyber.gouv.fr/en/publications/anssi-views-post-quantum-cryptography-transition}
\bibitem{r25} IETF Hackathon, pqc-certificates: Post-quantum cryptography certificates [software], n.d. \url{https://github.com/IETF-Hackathon/pqc-certificates} (accessed 11 August 2026).
\bibitem{r26} V. Dukhovni, Opportunistic Security: Some Protection Most of the Time, RFC 7435, IETF, 2014. \url{https://doi.org/10.17487/RFC7435}
\bibitem{r27} C. Evans, C. Palmer, R. Sleevi, Public Key Pinning Extension for HTTP, RFC 7469, IETF, 2015. \url{https://doi.org/10.17487/RFC7469}
\bibitem{r28} E. Rescorla, The Transport Layer Security (TLS) Protocol Version 1.3, RFC 8446, IETF, 2018. \url{https://doi.org/10.17487/RFC8446}
\end{thebibliography}
